\documentclass[useAMS,usenatbib,usegraphicx]{mn2e}
\usepackage{aas_macros}

\title[Satellite Luminosity Functions]{The
  Satellite Luminosity Functions of Galaxies in SDSS}
\author[Guo et al]{Quan Guo, Shaun Cole, Vincent Eke, Carlos Frenk\\
Institute for Computational Cosmology, Department of Physics, 
Durham University, Science
Laboratories, South Rd, Durham DH1 3LE.} 

\newcommand{\satmc}{M_{\rm{c}}}
\newcommand{\satdmbin}{\Delta M_{\rm{bin}}}
\newcommand{\satdmf}{\Delta M_{\rm{faint}}}
\newcommand{\satdzs}{\Delta z_{\rm{s}}}
\newcommand{\satap}{\alpha_{\rm{p}}}

\newcommand{\ri}{R_{\rm{inner}}}
\newcommand{\ro}{R_{\rm{outer}}}
\newcommand{\aii}{A^{\rm{inner}}_i}
\newcommand{\aio}{A^{\rm{outer}}_i}
\newcommand{\lii}{N^{\rm{inner}}_i(M)}
\newcommand{\lio}{N^{\rm{outer}}_i(M)}
\newcommand{\lsat}{N^{\rm{sat}}_i(M)}
\newcommand{\lsatj}{N^{\rm{sat}}_i(M_j)}
\newcommand{\mlsat}{\widetilde{N}^{\rm{sat}}(M_j)}
\newcommand{\satset}{\{\satmc,\ri,\ro, \satdmbin,\satdmf, \satdzs,\satap\}}
\newcommand{\satsets}{\{ \ri,\ro, \satdmbin,\satdmf, \satdzs,\satap\}}
\newcommand{\lcdm}{$\rm{\Lambda CDM}$}

\newcommand{\gnt}{N_{\rm T}}
\newcommand{\gns}{N_{\rm s}}
\newcommand{\gnbz}{N_{\rm b0}}

\newcommand{\gnbi}{N_{\rm b1}}
\newcommand{\gnbii}{N_{\rm b2}}
\newcommand{\gnbiii}{N_{\rm b3}}
\newcommand{\gnbv}{N_{{\rm b}i}}
\newcommand{\gnout}{N_{\rm out}}

\newcommand{\gmnt}{\langle \gnt \rangle}
\newcommand{\gmns}{\langle \gns \rangle}

\newcommand{\gmnout}{\langle \gnout \rangle}
\newcommand{\gmnbz}{\langle \gnbz \rangle}

\newcommand{\gmnbv}{\langle \gnbv \rangle}
\begin{document}

\pagerange{\pageref{firstpage}--\pageref{lastpage}} 

\maketitle

\label{firstpage}

\begin{abstract}
We study the luminosity function of satellite galaxies around isolated
primaries using the Sloan Digital Sky Survey (SDSS) spectroscopic and
photometric galaxy samples. We select isolated primaries from the
spectroscopic sample and search for potential satellites in the much
deeper photometric sample.  For primaries of similar luminosity to the
Milky Way and M31, we are able to stack as many as $\sim\negthinspace
20,000$ galaxy systems to obtain robust statistical results.  We
derive the satellite luminosity function extending almost 8 magnitudes
fainter than the primary galaxy. We also determine how the satellite
luminosity function varies with the luminosity, colour and
concentration of the primary. We find that, in the mean, isolated
primaries of comparable luminosity to the Milky Way and M31 contain
about a factor of two fewer satellites brighter than $M_V=-14$ than
the average of the Milky Way and M31.
\end{abstract}


\section{Introduction}
The \lcdm \ model predicts that structure forms in a hierarchical
manner.  Large spiral galaxies like the Milky Way (MW) and M31 form
within extended dark matter halos from the merging and accretion of
smaller subhalos. Smaller structures falling into bigger haloes can
survive there as substructures and host observed satellite galaxies.
A strong prediction of the theory, borne out by high resolution N-body
simulations, is that a very large number of such dark matter
substructures should survive in galactic halos \citep{kly99,moo99,
diemand07,springel08}. Empirical tests of this prediction have so far
been restricted to a single system, the Local Group, the only one for
which an estimate of the satellite luminosity function (LF) is readily
available.  Indeed, \citet{kly99} and
\citet{moo99} noted that the observed number of satellites around the
MW and M31 is much smaller than the number of predicted
substructures, giving rise to the so-called ``missing satellites
problem" of the \lcdm\ model.

In the past decade or so, fainter satellites around the MW 
and M31 have been discovered in the SDSS \citep[e.g.][]{bel08,bel10,
gre00,irw07,liu08,mar06,mar08,sim07,van00, wat09,zuc04,zuc06,zuc07},
but the number is still orders of magnitude smaller than the predicted
number of surviving cold dark matter subhalos. A number of solutions
to this problem have been proposed. Some invoke a different kind of
dark matter, warm dark matter, in which case the number of surviving
substructures is dramatically reduced by a cutoff in the primordial
power spectrum \citep{moo00, spe00,yos00,bod01,cra01,lovell11}. Others retain
cold dark matter and appeal to galaxy formation processes, such as
photoionization and supernova feedback, to inhibit star formation in
small subhalos thus rendering most of them invisible. This idea, first
mentioned nearly 20 years ago by \cite{kauffmann93}, was worked out in
detail a decade later using analytical arguments and semi-analytical
models \citep{bullock00,ben02,som02}.

The discovery of new Local Group satellites in the SDSS has stimulated
further observational and, particularly, theoretical work. \citet{kop08} and
\citet{tol08} extended the estimate of the satellite 
LF of the MW and M31 to faint magnitudes, accounting for the survey
magnitude limit and modelling the radial density profile of the
satellite distribution. This extension to faint magnitudes agrees
remarkably well with the \lcdm\ model predictions of
\cite{ben02}, a result that has been confirmed in recent
work using related semi-analytic modelling techniques
\citep{koposov09,munoz09,busha10,cooper10,mac10,li10,font11}. Full
N-body/gasdynamic simulations have also been carried out to
investigate the physics of satellite galaxies
(\citealt{libeskind07,okamotof09,oka10,wad10}) although currently
these simulations only resolve the brightest examples. In spite of
this broad agreement, interesting discrepancies exist. For example,
the original model of \citet{ben02}, as well as the more recent model
by \citet{qi10}, rarely produce satellites as bright as the LMC and
SMC \citep{boy10}.

The large body of work on satellite galaxies reflects the importance
of these objects as a critical test of the \lcdm\ model on small
scales. Yet, all conclusions to date regarding the validity or
otherwise of the model rely on comparison with data for a few dozen
satellites around just two galaxies, the MW and M31. There is no
guarantee that these are typical and indeed there is good evidence
that the satellites of the two galaxies have different structural
properties \citep{mcconnachie06,collins10}. Clearly, robust and
reliable tests of cosmological and galaxy formation models require
comparison with statistically representative samples of galaxies and
their satellites.
 
Analyzing the satellite sytems of external galaxies is challenging
because typically only one or two satellites are detected per primary
galaxy \citep{hol69, lor94,zar93,zar97b}.  In addition, the real space
position of the satellite with respect to its primary is uncertain. To
circumvent the first problem, these authors developed the method of
stacking the primaries and their satellites in order to obtain a
fair and complete sample which can yield statistically
robust results for certain classes of primary galaxies. These early
studies were limited by the relatively small samples available at the
time.  With the advent of large galaxy redshift surveys such as the
2dF Galaxy Redshift Survey
\citep[2dFGRS;][]{col01} and the Sloan Digital Sky Survey
\citep[SDSS;][]{yok00}, it is now possible to construct external galaxy
samples covering a much larger volume.  Studies with significantly
improved statistics have been carried out using these new surveys
\citep[e.g.][]{sal04,yan06,agu10},

In this work, we are interested in the satellite luminosity function
of specific types of isolated primary galaxies and, for this, the new
spectroscopic surveys are still not deep enough. For example, even
within the largest galaxy redshift catalogue from SDSS DR7
\citep{aba09}, where there are about $\sim \negthinspace 660\,000$
galaxies with $m_r^{\rm{lim}} < 17.77$, only a relatively small number
of isolated low redshift galaxy systems have enough detected
satellites for our purposes \citep[e.g.][]{hwa10}.  On the other hand,
the photometric catalogue from SDSS DR7 contains $\sim \negthinspace
96\,000\,000$ galaxies with magnitudes in the $u,g,r,i,z$ bands
(roughly $m_r^{\rm{lim}} \la 22$) and photometric redshifts. In this
study, we used both the spectroscopic and photometric SDSS DR7
catalogues. To ensure completeness, we restrict the photometric sample
in our main analysis to galaxies brighter than $m_r=20.5$ (see Section
4). The resulting catalogues enable us to analyze a sufficiently large
statistical sample of galaxy systems. We construct our sample using
methods similar to those developed by \cite{lor94} but modified
slightly to include photometric redshifts.

As this project was nearing completion, \cite{liu10} published an
investigation using similar methods to quantify the frequency with
which  satellites as bright as the LMC and
SMC occur around primaries similar to the MW. Shortly afterwards, 
\cite{lar10} also published a similar study, focused on
primaries brighter than $M_r<-20.5$, investigating how the satellite
LF and projected density profile depend on the primary luminosity and
colour. Our work complements these studies by including a wider range
of primary luminosities and exploring how the satellite LF depends on
the properties of the primary. Also, we adopt stricter isolation
criteria than those of \cite{lar10}. We compare our results with those
of these studies in the discussion in Section~\ref{sec:con}.

The remainder of this paper is organised as follows. In
Section~\ref{sec:data}, we describe the selection of primary galaxies
and their satellites; in Section~\ref{sec:method}, we develop the
method of estimating the satellite LF; in Section~\ref{sec:results},
we present our estimate of the satellite LF for different types of
primary galaxy. We conclude, in Section~\ref{sec:con}, with a summary
and discussion of our results.

\begin{figure*}
\begin{minipage}{126mm}
   \includegraphics[width=126mm]{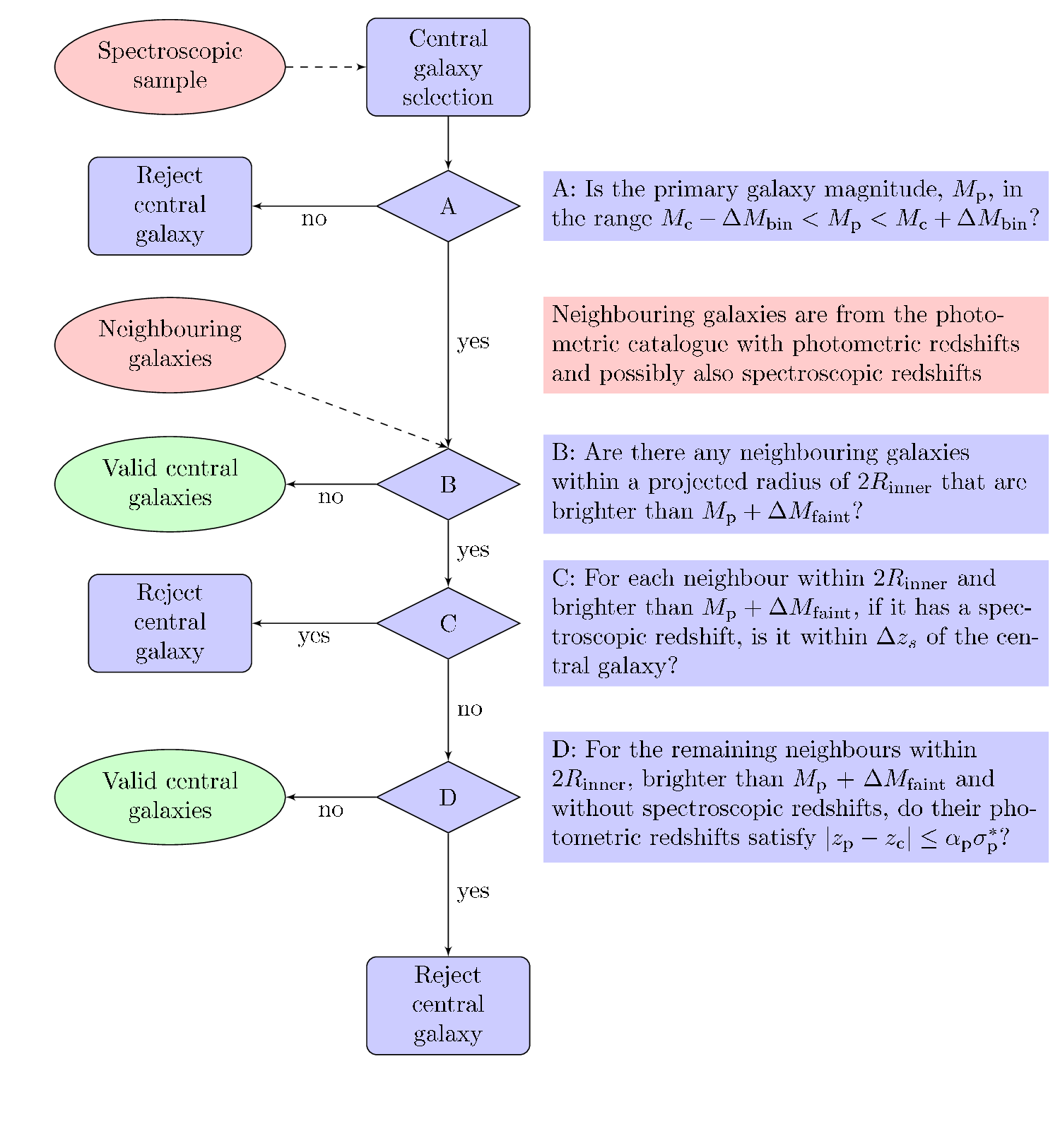}
   \caption{A flow chart detailing the selection criteria for isolated
     primary galaxies.}
   \label{fig:primary}
   \end{minipage}
 \end{figure*}

\section{Data and Sample Selection}
\label{sec:data}

We build two different catalogues for our study: a smaller one of
galaxies with spectroscopic redshifts from which we select the primary
galaxies (hereafter the spectroscopic catalogue) and a larger one of
galaxies with photometric redshifts and magnitudes from which we
select the neighbouring galaxies (hereafter the photometric
catalogue). The spectroscopic catalogue is constructed from the SDSS
DR7 spectroscopic subsample (north galactic cap) including all objects
with high quality redshifts (zconf $> 0.7$ and specClass $=2$) and a
Petrosian magnitude $r \le 17.77$.  The photometric galaxy catalogue
is from the SDSS DR7 photometric subsample (north galactic cap) and
includes only objects that have photometric redshifts, none of the
flags BRIGHT, SATURATED, or SATUR\_CENTER set and model magnitudes $r
\le 22.0$. We select only objects with corresponding entries in the
SDSS database PhotoZ table, which naturally selects galaxies and
excludes stars.  As galaxies with $r \le 17.77$ are included in both
SDSS catalogues, a small fraction of the photometric catalogue
galaxies also have spectroscopic redshifts.  We use de-reddened model
$ugriz$ magnitudes and k-correct all galaxies to $z=0$ with the IDL
code of \citet{bla07}. In addition, we also include $V$-band
magnitudes estimated from $g$ and $r$-band magnitudes assuming
$V=g-0.55(g-r)-0.03$ \citep{smi02}. This allows us directly to compare our results
with observations of the MW.

For our statistical analysis, the sample of primary galaxies should be
not only homogeneous but also isolated.  To this end, we adopt a series
of selection criteria summarised in the flow chart shown in
Fig.~\ref{fig:primary}.  First, from the spectroscopic sample, we
select primary galaxy candidates of absolute magnitude, $M_{\rm p}$, in
the range $M_{\rm{C}} -\Delta M_{\rm{bin}} < M_{\rm p} \le M_{\rm{C}}
+\Delta M_{\rm{bin}}$.  We then reject from this list those candidates
that have, or could have, bright neighbours whose own satellite system
could overlap with that of the candidate. We achieve this by rejecting
candidates that have a neighbouring galaxy within a projected distance
of $2\ri$ that is brighter than $M_{\rm p}+\Delta M_{\rm
  faint}$, unless that neighbouring galaxy is at a substantially
different redshift.  For neighbours with spectroscopic redshifts,
$z_{\rm s}$, the required redshift separation is $\vert z_{\rm p}-
z_{\rm s} \vert>\Delta z_s$, while for those with only photometric
redshifts, $z_{\rm phot}$, we require $\vert z_{\rm p} -z_{\rm phot}
\vert>\alpha_{\rm p} \sigma^*_p$.  Here $\sigma_{\rm p}^*$ is the
photometric redshift error that we adopt (see Section~\ref{sec:method})  and
$\alpha_{\rm p}$ is a tolerance, which we will vary.
The isolation criteria guarantee that there are no luminous
neighbouring galaxies that are projected within $2\ri$ of
the primary, unless these luminous neighbours are sufficiently far 
away from the primary and appear here due to a chance projection.
Using the photometric redshift information to identify and remove
true background and foreground galaxies significantly increases
the number of primary galaxies retained in our sample and reduces
the background contamination. 

\begin{table}
\caption{Properties of the primary galaxy samples for the following default
choices of values for the sample selection parameters, 
$\satset=\{\satmc, 0.3~\rm{Mpc}, 0.6~\rm{Mpc},0.5, 0.5, 0.002, 2.5\}$.
Quantities listed for each bin of $V$-band absolute magnitude $\satmc$ are:
the number of primary galaxy candidates (galaxies within the
absolute magnitude bin), the number of primary galaxies that pass
all the isolation criteria, their median redshift and redshift range.}
\label{tab:primary}
\begin{tabular}{@{}lllll}
\hline
$\satmc$ & primary & primaries & median & redshift  \\
      & candidates &\quad   &redshift &  range \\
\hline
-19.0 &  35893 & 88  & 0.043 & $0.021<z<0.066$ \\
-20.0 & 104907 & 2661&0.105 & $0.020<z<0.068$\\
-21.0 & 202351 & 21346 & 0.098& $0.016<z<0.164$\\
-22.0 & 94287 & 51733& 0.142&$0.022<z<0.391$\\
-23.0 & 51686 & 26982 & 0.203 & $0.031<z<0.522$\\
\hline
\end{tabular}
\end{table}

After having filtered by these criteria, the remaining isolated
galaxies comprise the primary galaxy catalogue. We briefly summarise
the properties of this catalogue. The number of primary galaxies not
only depends on their absolute magnitude, but also on the isolation
parameters.  The stricter the isolation criteria we take, the fewer
primary galaxies we have.  In the $V$-band, with a parameter set
$\satset=\{-21.0, 0.3~\rm{Mpc}, 0.6~\rm{Mpc},0.5, 0.5, 0.002,
2.5\}$\footnote{The parameter, $R_{\rm outer}$, is defined below},  the
number of candidates is 202\,351, which, after applying the isolation
criteria, is reduced to 21\,346 or about 10\% of the galaxies in this
magnitude bin. The primary galaxy redshifts lie in the range $0.01< z
< 0.16$, with a median redshift $0.098$. For different primary
magnitudes, $\satmc$, the number of primary galaxies and their median
redshifts are shown in Table~\ref{tab:primary}. For each magnitude
bin, the number of primary candidates is determined by the interplay
between the accessible volume given the survey limit and the density
of galaxies. The actual number of primaries is further affected by the
isolation criteria which, for example, tend to reject nearby galaxies
for which $2 R_{\rm inner}$ subtends a large angle.

\begin{figure*}
   \includegraphics[width=126mm]{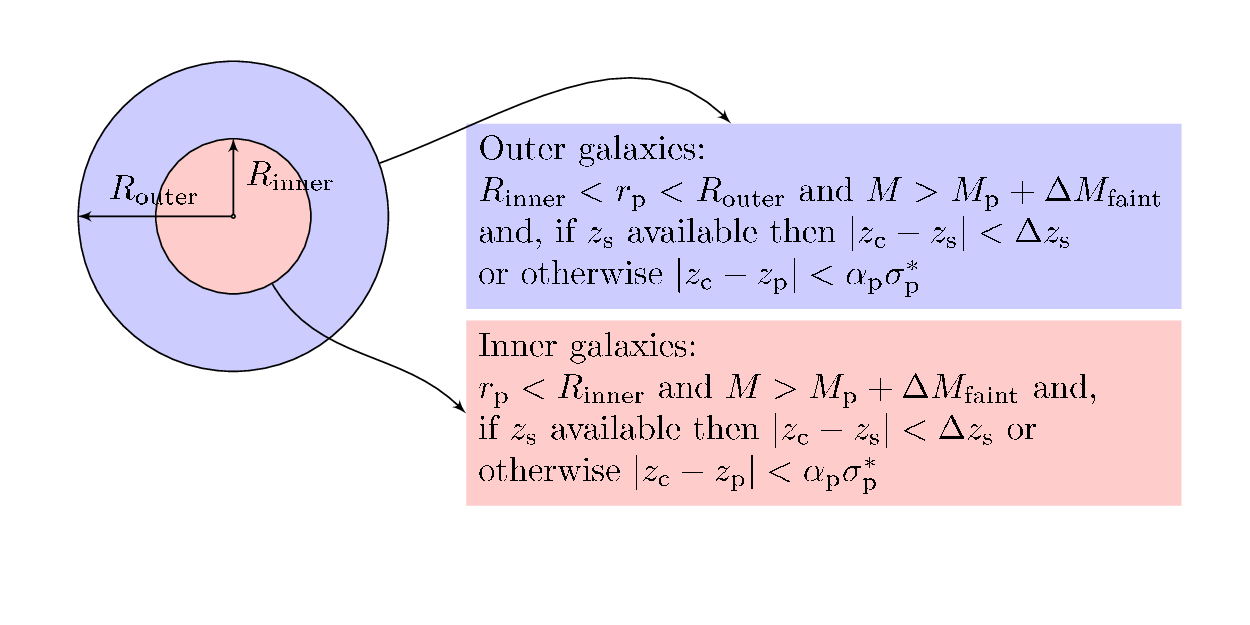}
   \caption{Schematic showing the selection of potential satellite galaxies
     within $\ri$ and of a reference sample within an annulus defined
     by $\ri<r<R_{\rm outer}$, used to subtract the residual
     contaminating background. For both samples we apply the stated
     redshift cuts to reduce background contamination. We also apply
     the stated absolute magnitude cut to both samples (assuming the
     neighbouring galaxies are at the same redshift as the primary)
     though this cut is redundant unless
     $R_{\rm{outer}}>2\ri$ as otherwise the existence of
     such bright neighbouring galaxies would automatically lead to the
     exclusion of the primary galaxy. }
   \label{fig:sat}
\end{figure*}

The schematic in Fig.~\ref{fig:sat} indicates our selection procedure
for potential satellites or ``inner galaxies'', and the corresponding
selection of the ``outer galaxies'' used to define the background.  We
assume the satellites of the primary galaxy fall within a projected
radius, $\ri$ (the red circle in Fig.~\ref{fig:sat}).  To reduce the
background contamination, we apply the same cuts in redshift
(spectroscopic and photometric) as were applied when selecting the
primary galaxies, but as most of the galaxies within $\ri$ only have
photometric redshifts with quite large measurement errors, we still
cannot distinguish true satellites from projected background galaxies.
However, the existence of satellites will make the number density of
galaxies within $\ri$ slightly larger than that in the outer blue
reference annulus in Fig.~\ref{fig:sat} ($\ri <r<R_{\rm{outer}}$).  By
counting the difference between the number density of galaxies within
$\ri$ and in the reference annulus, we can estimate the number of true
satellites.

\begin{figure*}
\begin{minipage}{168mm}
    \includegraphics[width=168mm]{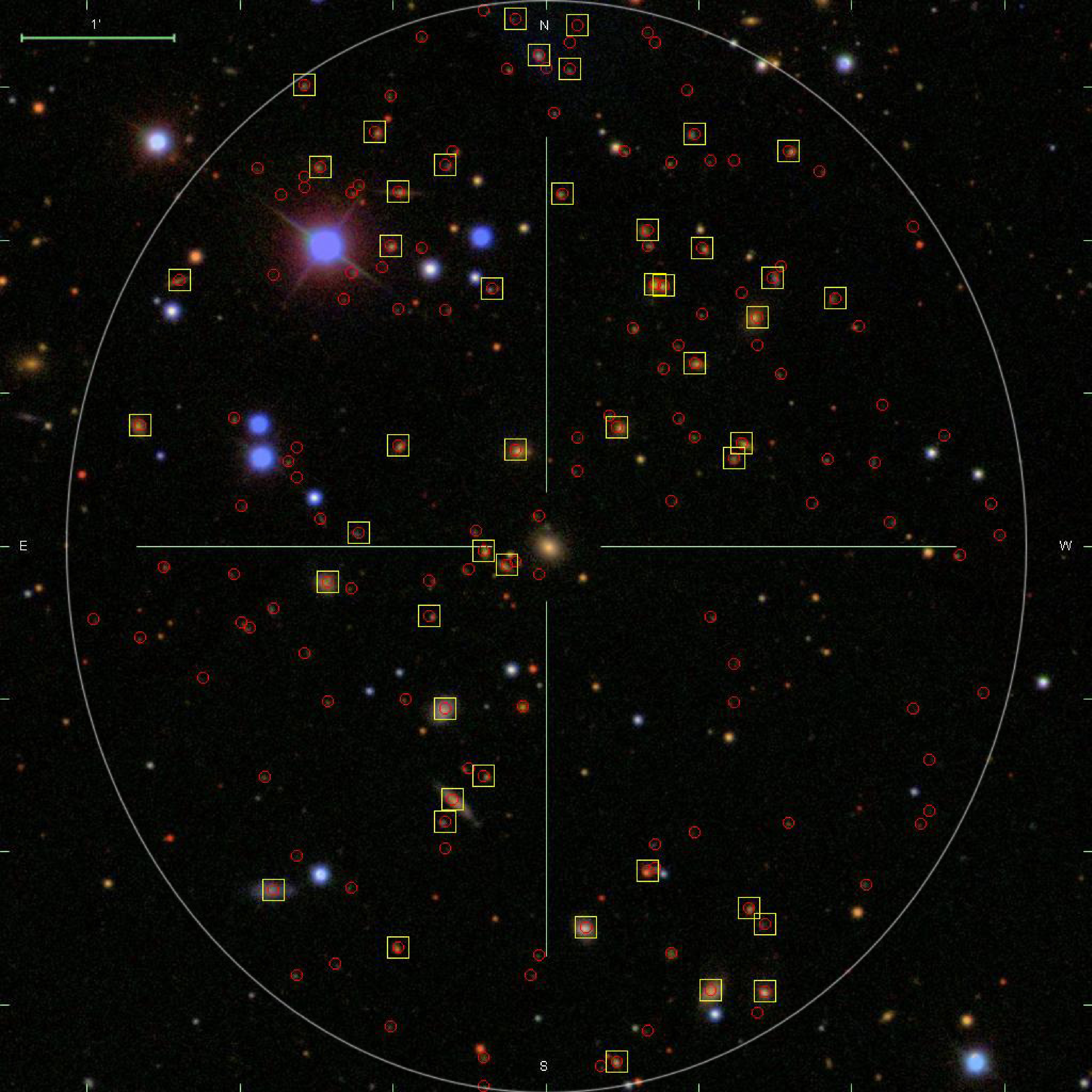}
   \caption{An example SDSS DR7 image centred on a primary galaxy of
     magnitude $m_r=16.10$ at redshift $z=0.074$. The white ellipse marks
     $r=\ri=300$~kpc.  All catalogued galaxies
   projected within $\ri$ of the primary are marked with red
   circles.  Those brighter than our fiducial $m_r=20.5$ magnitude
   limit are marked with yellow boxes.
   The remaining unmarked images within $\ri$ are presumed to be 
   classified as stars, which we have verified in this case using
   the manual    SDSS DR7 Navigate Tool.   
}
   \label{fig:min_max_sat}
   \end{minipage}
 \end{figure*}

An example of the objects we detect around a typical primary galaxy
is shown in Fig.~\ref{fig:min_max_sat}. This image, produced by the
SDSS finding chart
tool\footnote{http://cas.sdss.org/dr7/en/tools/chart/chart.asp},
illustrates the quality of the data and shows that candidate
satellites are spatially well separated from the light distribution of
the primary galaxy. The white circle (slightly stretched in this
Aitoff projection) indicates $r=\ri$. Within this region we have
marked all the galaxies in our catalogue with red circles and the
subset brighter than $m_r=20.5$, used in our main analysis, with
yellow boxes.  The remaining visible objects within $\ri$ are not in
our catalogue. Manual inspection with the DR7 Navigate tool reveals
them to be classified as stars.

\begin{figure}
\includegraphics[width=84mm]{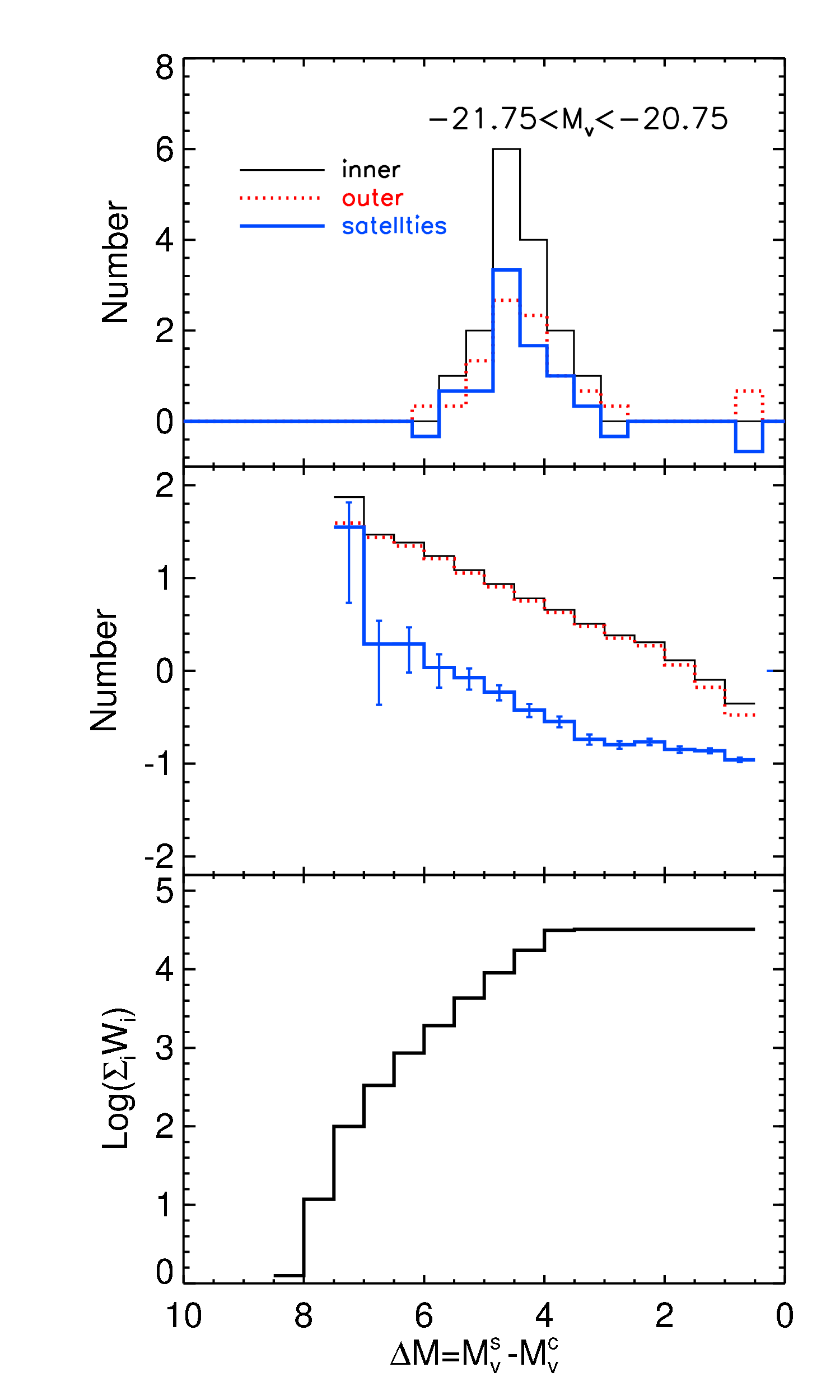}
\caption{Estimation of the satellite luminosity function. The top
panel shows the $V$-band LF for a single primary galaxy. The middle
panel shows the mean satellite LF of all primary galaxies. The black
(thin) and red (dotted) lines give the counts of inner and outer
galaxies respectively and the blue (thick) lines the estimate of the
satellite LF. The number of primary galaxies contributing to the mean
satellite LF in each bin is shown in the bottom panel.  Here the
selection parameters, $\satsets$, are set to the default values
$\{-21.25, 0.3~\rm{Mpc},0.6~\rm{Mpc},0.5, 0.5, 0.002, 2.5\}$}
\label{fig:est_pry}
\end{figure}

\section{Estimating the Satellite Luminosity Function}
\label{sec:method}
Once the primary galaxies are defined, their potential satellites
are found from the photometric galaxy
catalogue as depicted in Fig.~\ref{fig:sat}.
For the $i$th primary galaxy, the number of inner
galaxies, $\lii$, is found by counting all neighbouring galaxies
within the inner area that satisfy the following conditions: at least $\satdmf$
fainter than the primary; if they have a spectroscopic redshift, 
$z_{\rm{s}}$, then it should satisfy $|z_{\rm{c}}-z_{\rm{s}}|<\satdzs$;
or if they only have a photometric redshift $z_{\rm{p}}$, then it
should satisfy $|z_{\rm{c}}-z_{\rm{p}}|<\alpha_{\rm{p}}\sigma_{\rm{p}}^*$, where
$\sigma_{\rm{p}}^*$ is the error  in the photometric redshift as
defined below. The number of outer galaxies, $\lio$, is
determined by applying the same conditions to galaxies in the outer
area. As most
satellites of the primary should be projected within $\ri$ of the
primary, the number density of inner galaxies should typically exceed
that of the outer galaxies.  The excess can be taken as the projected
satellite LF of the $i$th primary galaxy, and estimated by
\begin{equation}
\lsat=\lii-\frac{\aii}{\aio}\,\lio,
\end{equation}
where $\aii$ and $\aio$ are the areas of the inner and outer regions
respectively (excluding sub-regions not within the sky coverage of
the SDSS DR7, which we have identified using the mask described in
\citet{norb11}) .

Because of the survey apparent magnitude limit, we are able to probe less
of the faint end of the satellite LF for primaries at higher redshift.
To account for this and construct an
unbiased estimate of the satellite LF averaged over all primary
galaxies, we count the effective number of primaries contributing to
each bin of the LF using the weighting function
\begin{equation}
W_{ij}(M_j)=\left\{
\begin{array}{ll}
1 &  M_j < M_i^{\rm{lim}}+\Delta  M_j \\
\frac{(M_i^{\rm{lim}}-\Delta M_j-M_j)}{2\Delta M_j} 
&M_i^{\rm{lim}}-
\Delta M_j\le  M_j \le \\
 & \qquad\qquad M_i^{\rm{lim}}+\Delta M_j  \\
0 &   M_j > M_i^{\rm{lim}}-\Delta M_j
\end{array}\right. ,
\end{equation}
where $ M_j$ is the central value of each magnitude bin, $\Delta M_j$
is the half width of the bin,
$M_i^{\rm{lim}}=m^{\rm{lim}}-5\log_{10}(D^L_i)-K(z_i)$, $D^L_i$ is the
luminosity distance of the i$th$ galaxy and $m^{\rm{lim}}$ is the SDSS
galaxy spectroscopic sample magnitude limit. For a given primary, the
weighting function is unity for all magnitude bins in which satellites
anywhere in the bin are bright enough to be included in the survey. It
is zero if all satellites within the bin are too faint to be included
in the survey and ramps between zero and one when only galaxies in a
fraction of the bin width are accessible to the survey.  We then
define the effective number of primary galaxies, $N^{\rm {prim}}_j$, 
contributing to the $j$th bin of the LF as $N^{\rm {prim}}_j = \sum_i
W_{ij}(M_j)$. With this definition, our unbiased estimator of the
average satellite LF is given by
\begin{equation}
{\mlsat}=\frac{\sum_i \lsatj }{N^{\rm prim}_j} .
\label{eqn:est}
\end{equation} 

In practice, in our study we divide the satellite luminosities, $M_j$,
into $20$ bins ($j=1,2,\cdots,20$). Furthermore, because each primary
galaxy in the same bin has a slightly different magnitude relative to
$\satmc$, we choose to show our results in terms of the difference
in the magnitude of the satellite and primary galaxy, $\Delta M=
M_{\rm{s}}-M_{\rm{p}}$, which aligns the satellite LFs in
the same bin.

The process of estimating the satellite LF for primaries in one bin of
$V$-band absolute magnitude is illustrated in
Fig.~\ref{fig:est_pry}. The thin black histogram in the top panel
shows the number of inner galaxies binned by $V$-band magnitude
difference for one of the primaries.  The dotted red histogram shows
the corresponding number of outer galaxies scaled by the ratio of
areas ${\aii}/{\aio}$. Their difference, which is an estimate of the
satellite LF in that system, is shown by the thick blue histogram.
The thin black and dotted red histograms in the middle panel show the
number of inner and (scaled) outer galaxies per primary where the
number of primaries, $N^{\rm prim}_j=\sum_i W_{ij}(M_j)$, contributing
at each $\Delta M$ is shown in the lower panel. The heavy blue
histogram in the middle panel of Fig.~\ref{fig:est_pry} shows the
estimated mean satellite LF for all primaries in the magnitude range
$-21.75<M_{\rm V}<-20.75$.  The error bars on this mean satellite LF
are estimated by bootstrap resampling of the set of primaries. At the
faint end of the LF the error bars become quite large because of the
small number of nearby primaries that are able to contribute. If the
faintest bin only contains one primary then we show a Poisson, rather
than the bootstrap error.

\begin{figure*}
\includegraphics[width=148mm]{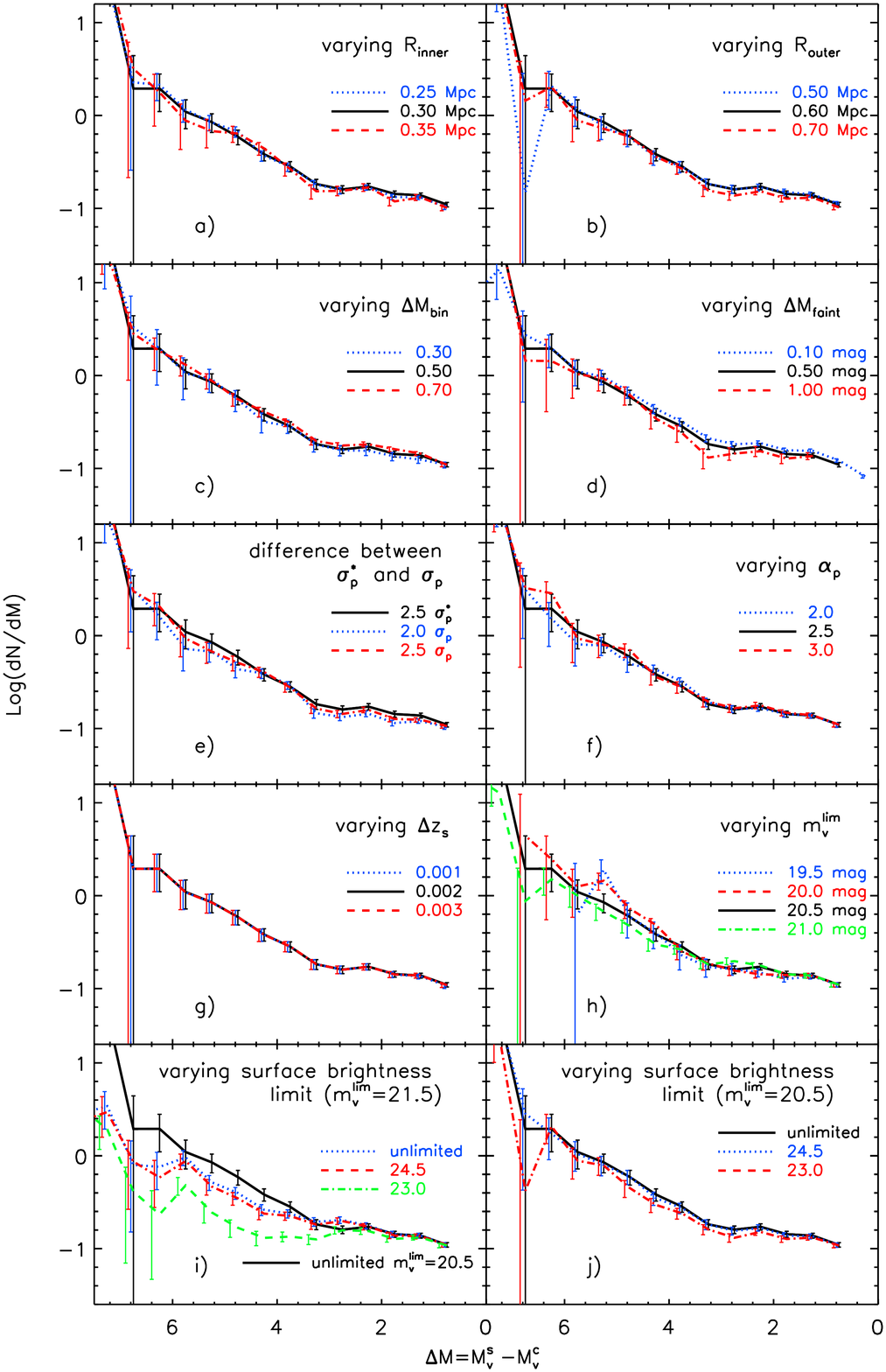} 
\caption{The effect on the estimated satellite LF of varying the
  parameters $\satsets$ from their default values, $\{-21.25,
  0.3~\rm{Mpc},0.6~\rm{Mpc},0.5, 0.5, 0.002, 2.5\}$, as indicated
  in the legends. In addition, panel~e shows the effect of 
  changing the assumed photometric redshift error from the original 
  $\sigma_{\rm p}$ to our adopted $\sigma_{\rm p}^*=\max(\sigma_{\rm p},0.05)$ .
  Panels~h,i and~j show the effect of varying the apparent 
  magnitude limit of the photometric catalogue and imposing an additional
  cut on surface brightness (see text for details). In this and in
Figs.~6, 7 and 9, the error bars for different datasets have been
slightly shifted for clarity.
}
\label{fig:varyingpars}
\end{figure*}

For a specific $\satmc$, the selection of primaries and counts of inner
and outer galaxies are determined by the parameter set $\satsets$. 
It is important to choose appropriate values for these parameters. 
Here we discuss the physical motivation for our choice of parameter
values and check that the resulting satellite luminosity function is robust to
reasonable variations in these parameters. The various panels in 
Fig.~\ref{fig:varyingpars} show the results of varying these
parameters away from our default choice of
$\satset=\{\satmc, 0.3~\rm{Mpc}, 0.6~\rm{Mpc},0.5, 0.5, 0.002, 2.5\}$.

The area within which we search for the satellite signal is determined
by the parameter $\ri$. For too small a value of $\ri$, we would lose
genuine satellites. Once $\ri$ is sufficiently large to enclose all
the true satellites the resulting background-subtracted satellite LF
should be independent of $\ri$. However, the statistical error in the
estimate will increase due to increased background contamination.  The
value of $0.3~\rm{Mpc}$ is roughly the virial radius of the Milky Way,
and so this seems a reasonable value to take for the $\ri$ of Milky
Way-like primary galaxies. One could argue for scaling $\ri$ with the
magnitude or type of the primary galaxy, but, for simplicity, we set
$\ri=0.3~\rm{Mpc}$ in this study except in our parameter tests.  In
Fig.~\ref{fig:varyingpars}a, we show that the effect of varying $\ri$
between $0.25$ and $0.35$~Mpc does not change the satellite LF
significantly. A possible concern is that the SDSS data reduction
pipeline occasionally misclassifies fragments of the spiral arms of
bright galaxies as separate galaxies. We have checked that these
contaminating objects do not make a significant contribution to our
estimate of the satellite luminosity by excluding all galaxies within
$1.5$ times the Petrosian $R_{90}$ radius of the primary
galaxies. Comparison of the resulting satellite luminosity functions
shows that they make no significant difference.

The next parameter, $\ro$, determines the outer reference annulus from
which we estimate ``background" counts. An appropriate value for $\ro$
will guarantee a suitably local estimate of the background. A local
estimate of the background is preferable \citep[see][]{chen06} as
galaxies are clustered and, in our case, the mean environment of a primary
galaxy is also biased by the isolation criteria that we apply.
Fig.~\ref{fig:varyingpars}b shows that, provided the outer area is
sufficiently large to allow an accurate estimate of the background,
the resulting satellite LF is robust to changes in $\ro$. 
We also tested the effect of estimating the background using a larger
annulus that was disjoint from the inner region 
(from $0.5$~Mpc to $0.7$~Mpc ) and again found no significant
difference.

Besides the physically motivated parameters, we also test the
parameters of the estimation method.  For a specific central
magnitude, $\satmc$, the bin half width, $\satdmbin$, is a
compromise between having a large enough sample of primary galaxies
and not distorting the LF due to averaging over primaries of differing
luminosities.  Fig.~\ref{fig:varyingpars}c shows results for a few
different $\satdmbin$ values and indicates that, for our choice of
binning, the satellite LF by the magnitude difference, $\Delta M=
M_{\rm s}-M_{\rm p}$, any biases are very small.

The next panel, Fig.~\ref{fig:varyingpars}d, shows the effect of
varying the parameter $\satdmf$, which is important in selecting
isolated primaries. The larger $\satdmf$, the smaller the number of
primary galaxies that survive the isolation filter. Hence, the value of
$\satdmf$ is a compromise between avoiding the introduction of
primary galaxies within groups and
gathering sufficient primary galaxies. We adopt $\satdmf=0.5$, but
Fig.~\ref{fig:varyingpars}d shows that, apart from the truncation of
the satellite LF brighter than $\Delta M =\satdmf$, the results are,
perhaps surprisingly, insensitive to changing to $\satdmf=0.1$ or
$1.0$. To test further the effect of varying the isolation criteria
we have cross matched our primary galaxy catalogue with the 
\citet{yan07} group catalogue. We find that within the DR4 footprint
of the Yang et al catalogue only 467 of our $\sim \negthinspace 20\,000$ primary
galaxies for our fiducial value of $M_c$ and $\Delta M_{\rm
faint}=0.5$ match with groups of 2 or
more galaxies. Excluding these group members from our list of
primaries has essentially no effect on the estimated LF and so we
conclude that our satellite LF has no significant contamination from
group members.

The parameter $\satap$ helps us to distinguish genuine satellite
galaxies from background galaxies by excluding galaxies that are at a
significantly different redshift. If too small a value of $\satap$ is
used then we will artificially exclude genuine satellite galaxies just
because the random error in their photometric redshift happens to be
greater than $\satap\sigma_{\rm p}$.  If the quoted $\sigma_{\rm p}$ were accurate
for all galaxies and the errors were Gaussian then $\satap>2$ ought to
be sufficient. However the dotted blue and dashed red lines in
Fig.~\ref{fig:varyingpars}e show that with both $\satap=2$ and~$2.5$
the satellite LF is systematically underestimated at the faint
end. Further investigation has revealed that the cause of the
sensitivity is that some galaxies with low values of $\sigma_{\rm p}$ in
reality have larger redshift errors due either to non-Gaussian
distributions or inaccuracies in $\sigma_{\rm p}$. Hence, for our default
selection we have been more conservative and set a floor on the
photometric redshift error by adopting
$\sigma_{\rm p}^*=\rm{max}(\sigma_{\rm p},0.05)$.  Fig.~\ref{fig:varyingpars}f
shows that with this choice the satellite LF does not depend
systematically on $\satap$.

Fig.~\ref{fig:varyingpars}g shows the dependence of the satellite LF
on $\satdzs$, the maximum allowed
spectroscopic redshift difference between a satellite and its primary.
This should be large enough so that satellites are not excluded due to
the line-of-sight component of their orbital velocities.
Our default choice is $\satdzs=0.002$, corresponding to a line-of-sight velocity
difference of $600~\rm{km~s^{-1}}$.  The results are very insensitive
to this value, mainly as only a small fraction of our potential
satellites from the photometric catalogue have spectroscopic redshifts.

The final three panels of Fig.~\ref{fig:varyingpars} illustrate the
sensitivity of our results to the apparent magnitude and surface
brightness cuts that we impose on the photometric catalogue.
Fig.~\ref{fig:varyingpars}h shows that the satellite LF is
systematically suppressed at the faint end if all catalogued galaxies
are used to a faint magnitude limit of $m_V=21.5$, compared to our
default of $20.5$. Brighter cuts also cause some variation but in this
case the samples are becoming smaller and
noisier. Figs.~\ref{fig:varyingpars}i and~j show the effect of
applying cuts in surface brightness (mean surface brightness within
the Petrosian $R_{50}$ radius) for two different apparent magnitude
limits. For a faint magnitude limit of $m_V=21.5$ the faint end of the
LF is very sensitive to the surface brightness cut. This occurs
because the catalogue is not complete to $m_V=21.5$ and preferentially
misses low surface brightness galaxies. With the brighter default cut
of $m_V=20.5$ this effect is greatly reduced
(Fig.~\ref{fig:varyingpars}j), indicating much higher completeness and
little sensitivity to the surface brightness cut.  For the $r$-band
catalogue, we perform similar tests and find that cuts at similar
values to those found for the $V$-band are appropriate.

\begin{figure}

\includegraphics[width=84mm]{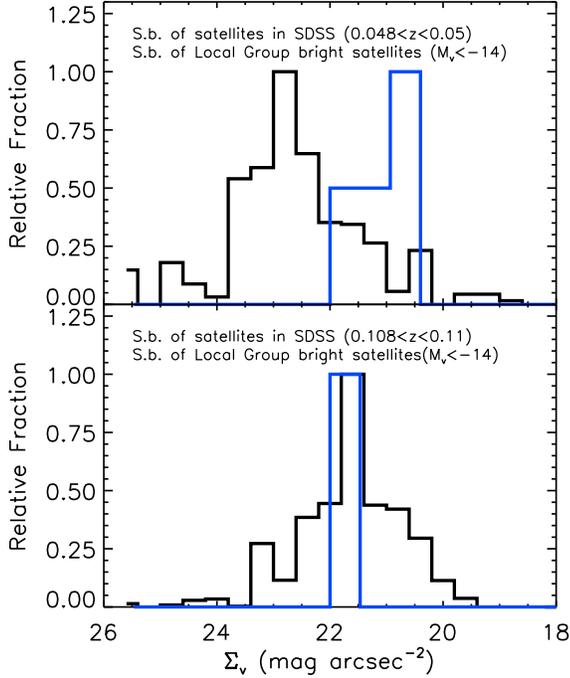}

\caption{A test of surface brightness effects. The black histograms
show the distribution of surface brightnesses, defined as the average
surface brightness within the half light radius, for potential
satellite galaxies around primaries at redshift $z\approx 0.05$ (upper
panel) and $z\approx 0.1$ (lower panel).  These are compared with the
surface brightness distribution (blue histograms) of those bright 
Local Group satellites which would be brighter than our $m_V=20.5$ apparent
magnitude limit when placed at the redshift of the selected
primaries. These equivalent surface brightnesses, computed from the
data provided by \citet{mat98}, have been k-corrected and redshift-dimmed
to the redshift of the selected primaries.}
\label{fig:surf}
\end{figure}

Some of the known Local Group satellites have quite low surface brightnesses
\citep{mat98} and it is important to check that their counterparts would not be
missed in our analysis by falling below the SDSS detection limit. In
Fig.~\ref{fig:surf} we plot the distribution of observed surface brightnesses
of galaxies around primaries in two different redshift intervals. The
turnover in these distributions at around $\Sigma_V\approx 23$ is to
be expected given the intrinsic distribution of galaxy surface
brightnesses \citep{dri05}. The distributions for the SDSS
spectroscopic survey only become incomplete around 
$\Sigma_V\approx 24\ {\rm mag\ arcsec^{-2}}$ \citep{str02}. The surface
brightness distributions of the subset of Local Group satellites whose absolute
magnitudes are sufficiently bright for them to be selected in our
catalogue are shown by the blue histograms.  These can be seen to have
surface brightnesses that fall near the middle of the measured
distribution. 

If the 8 Local Group satellites
considered for this study were gradually moved to higher redshifts, then
only NGC205 would drop out of our sample by having a surface brightness below 
$\Sigma_V=24\ {\rm mag\ arcsec^{-2}}$ before it was lost beneath the
flux limit.
As our sample also includes the SDSS DR7 photometric
subsample, we do actually detect satellites at surface brightnesses
below that of NGC205, so a conservative estimate of
the incompleteness due to low surface brightness is 1 in 8. M32 is
such a centrally concentrated satellite that 
it would be classified as a star by SDSS,
so there is also likely to be a comparably small incompleteness
at high surface brightness in our analysis.

These combined results show that our method of estimating the
satellite LF is quite robust to changes in the parameter values used
in the estimation method. Therefore we will use
$\satsets=\{0.3~\rm{Mpc}, 0.6~\rm{Mpc}, 0.5, 0.5, 0.002, 2.5\}$ in the
rest of the paper.

\begin{figure}

  \includegraphics[width=84mm]{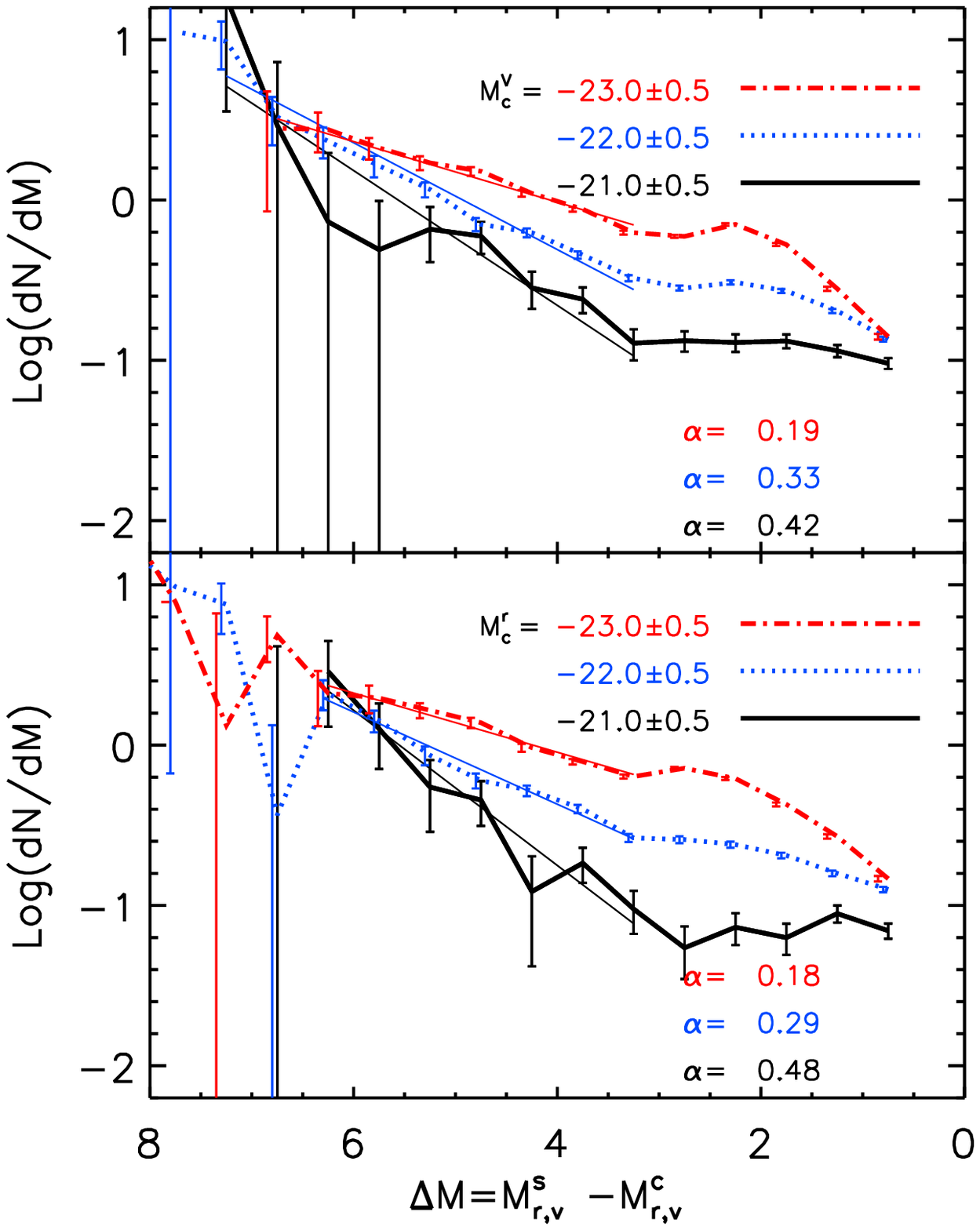} \caption{The estimated satellite
  LFs for different bins of primary magnitude as indicated in the
  legend. The top panel is for the $V$-band and the bottom panel for
  the $r$-band. The straight lines show power-law fits to the faint
  ends of the luminosity functions.  Their slopes, $\alpha$, are given
  in the legend.  Here, the selection parameters, $\satsets$, are set
  to the default values $\{ 0.3~\rm{Mpc},0.6~\rm{Mpc},0.5, 0.5, 0.002,
  2.5\}$}. \label{fig:diff_mc}
\end{figure}

\begin{figure}
\includegraphics[width=84mm]{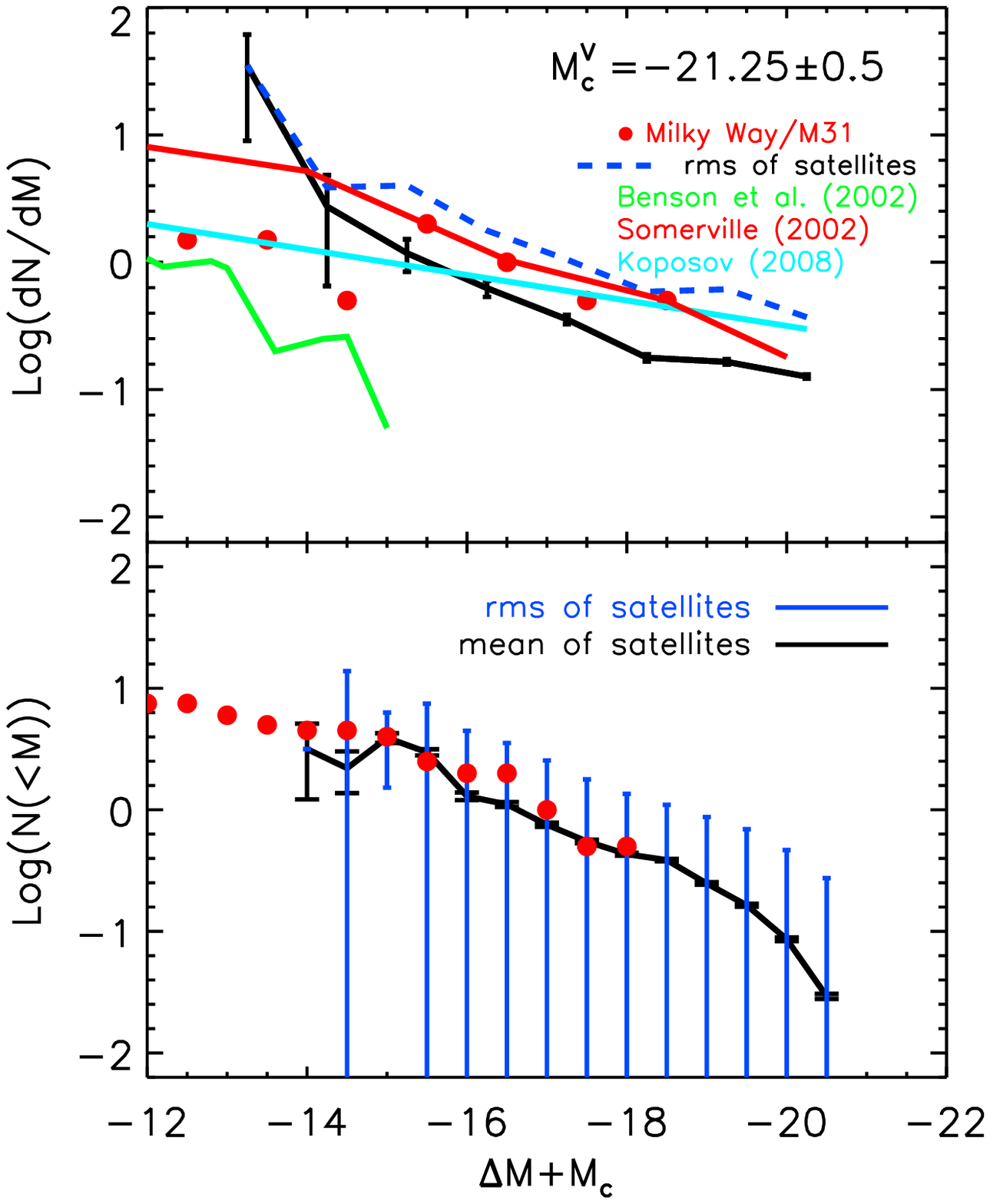}
   \caption{A comparison of the average satellite LF in our sample
   with the satellite LF of the Milky Way and M31. The upper panel
   shows the differential satellite LF of MW-like galaxies. The solid
   line with error bars shows the estimated $V$-band satellite LF of
   primaries with similar magnitudes to the Milky-Way and M31 ($\satmc
   = -21.25\pm 0.5$ in the $V$-band). This is compared to the mean LF
   of the MW and M31 (per central galaxy) in unit magnitude bins shown
   by the red points. The best fit power law,
   $dN/dM_v=10\times10^{0.1(M_v+5)}$, of \citet{kop08} is shown as the
   cyan line.  The theoretical predictions of \citet{ben02} and
   \citet{som02} for $z_{\rm{reion}}=10$ are shown by the green and
   red lines respectively.  The blue dashed line labelled ``rms of
   satellites'' shows the mean value plus the rms of the
   LF among different primaries. The lower panel shows the same
   results and the observational data in cumulative form. Here, the
   black error bars give the error on the mean cumulative LF while the
   much broader blue error bars indicate the intrinsic rms
   scatter about this mean.  } \label{fig:mw_lf}
\end{figure}

\section{Results}
\label{sec:results}

We now explore the dependence of the satellite LF on the properties of
the primary galaxies. Estimates of the V and $r$-band satellite LF for
primaries of magnitude $\satmc=-21, -22$ and~$-23$ are shown in
Fig.~\ref{fig:diff_mc}. As the luminosity of the primary increases the
number of satellites increases at all values of $\Delta M$, and, in
addition, the shape of the LF changes.  None of the luminosity
functions are well fit by Schechter functions, i.e. they are not well
described by power laws with exponential cutoffs at the bright
end.  Instead, there is a tendency for the LFs to become flatter at the
bright end and the satellite LFs of the brightest primaries even have
a local maximum at $\Delta M=2$. Only at the faint end are the
luminosity functions accurately represented by power laws. We show
such fits and list their slopes in Fig.~\ref{fig:diff_mc} .
The variety of features in the LFs
suggests they will place interesting constraints on formation models.

\begin{figure}
     \includegraphics[width=84mm]{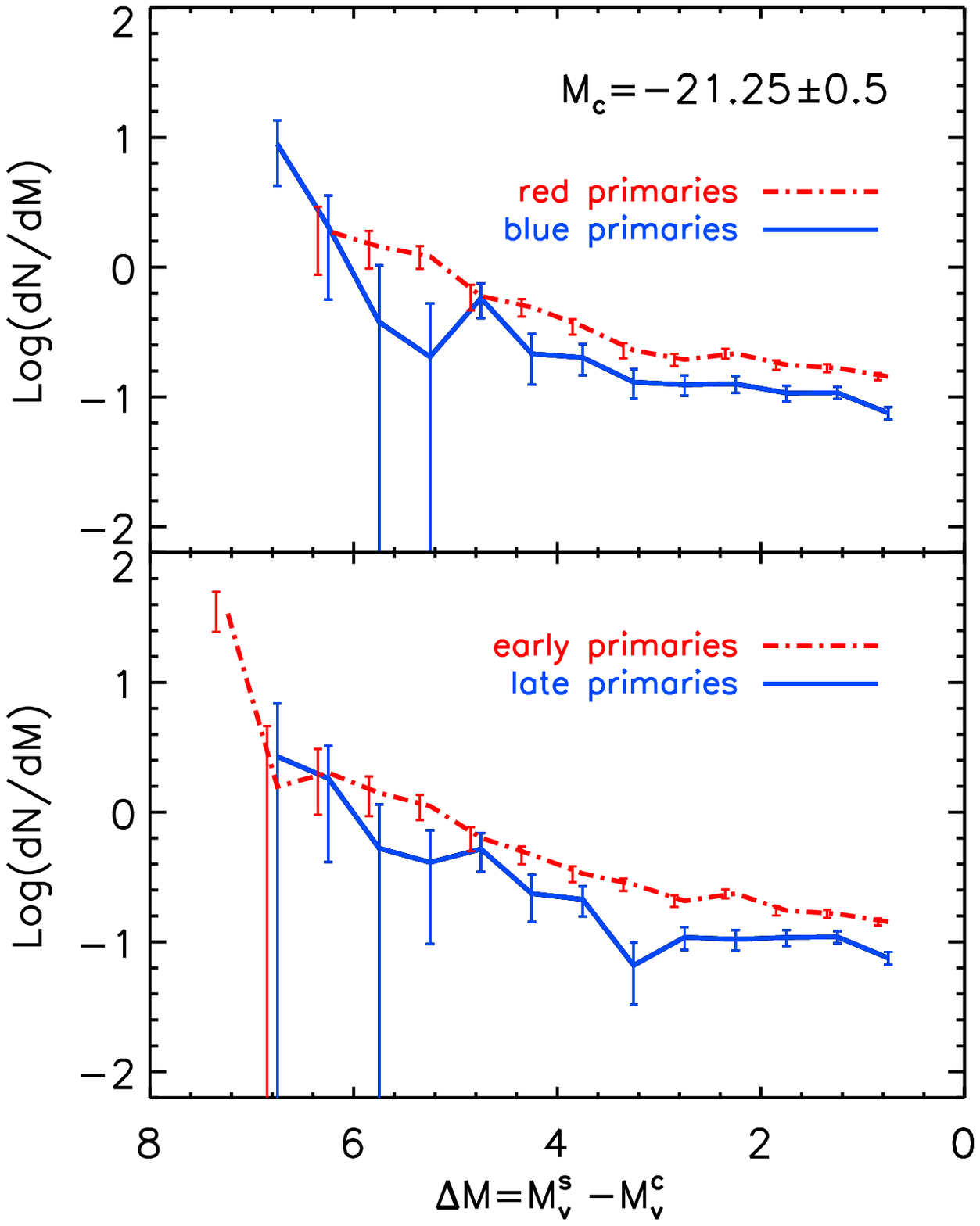}
   \caption{The mean satellite LF of different colours (top panel) and
     types (bottom panel) of primary galaxy. The satellite LF of
     early-type or red primary galaxies is shown as a red (dot-dashed)
     line and that of late-type or blue is plotted as a blue (solid)
     line.  }
   \label{fig:type_lf}
\end{figure} 

\begin{figure}
\includegraphics[width=84mm]{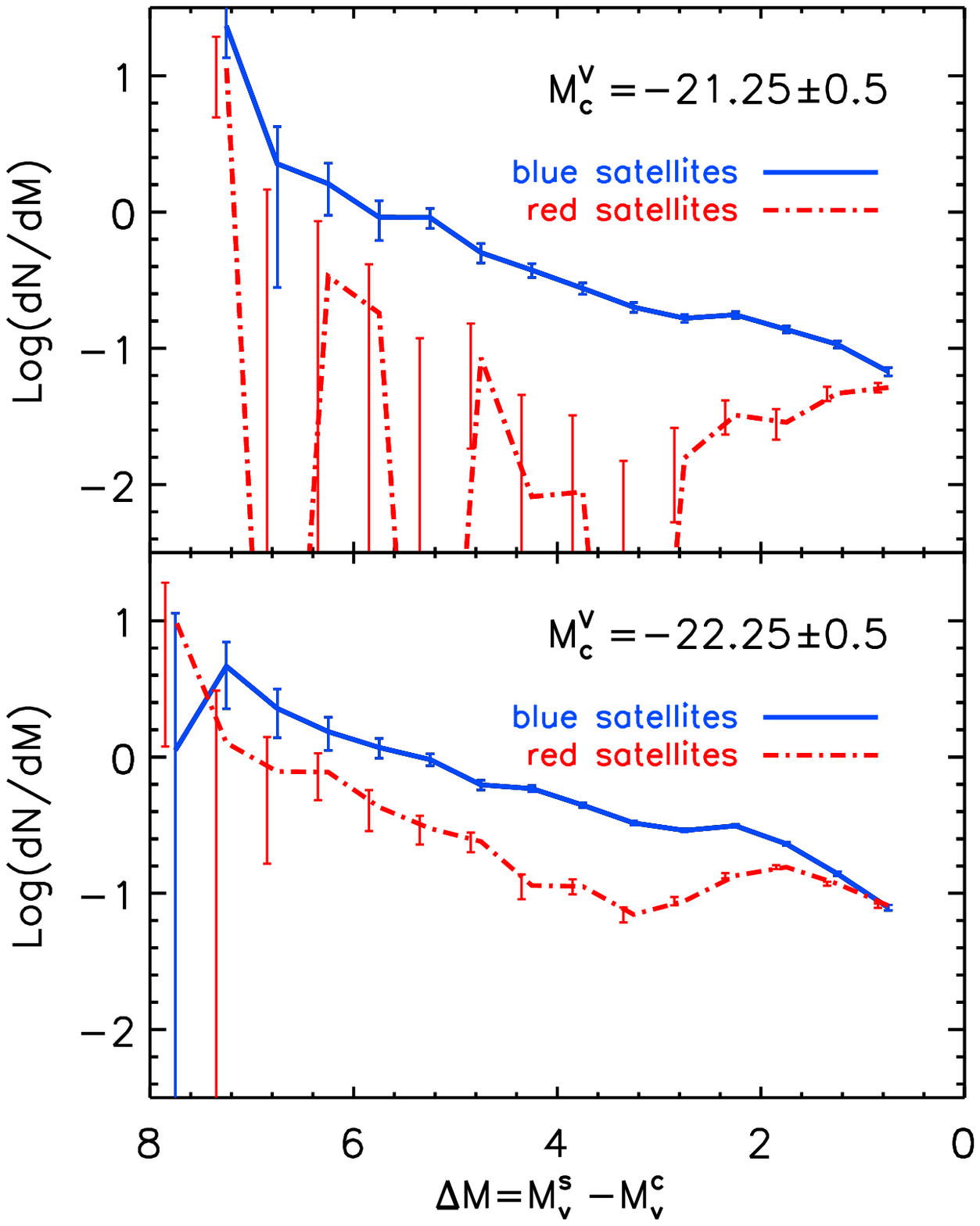} 
\caption{The satellite LF split into contributions from
  ``blue'' and ``red'' satellites.
  The top and bottom panels show the results for primaries of $V$-band
  magnitude $\satmc=-21.25$ and~$-22.25$ respectively. 
  The ``blue" and ``red'' satellite LFs
  are plotted as blue (solid) and red (dot-dashed)
  lines respectively. }
\label{fig:sat_color}
\end{figure}


In Fig.~\ref{fig:mw_lf}, we carry out a comparison of the satellite LF
of primaries of similar luminosity to the MW and M31 with data for
these two galaxies. The average satellite LF of the MW and M31 has
often been compared to theoretical models
\citep[e.g.][]{ben02,som02} and used to constrain properties of the
model such as the redshift of reionization and the strength of
supernova feedback. In so doing one implicitly assumes that the
satellite LF per primary galaxy of the combined MW+M31 system is
typical of isolated galaxies of similar luminosity. The data allow a
direct test of this assumption at the bright end, $M_{\rm V}<-14$, of
the LF. For this comparison, we assume that the $V$-band magnitudes of
both the MW and M31 lie in the range $-21.25\pm0.5$
\citep{fly06,gil07} and compare directly with the average of their
$V$-band LFs by plotting on the $x$-axis the $V$-band $\Delta M +
M_{\rm c}$. Over the range $-14>M_{\rm V}>-19$ our mean LF has a very
similar slope to that of the average of the MW and M31, but with
almost a factor two fewer satellites at all luminosities. Fainter than
$M_{\rm V}=-14$ our estimate becomes noisy due to a lack of nearby
primaries. The random errors on our estimate of the mean luminosity
density are small at bright magnitudes, yielding a well-defined
estimate of the luminosity function that provides a very strong
constraint on models all the way to magnitudes as bright as $M_{\rm
V}=-20$. Comparison with the theoretical models of
\citet{ben02} and \citet{som02} highlights the range of predictions.
Tuning the models to match our new data rather than just the MW or M31
may lead to a different assessment of the strength of feedback effects
in suppressing the formation of satellite galaxies. This is
particularly apparent when one considers the system-to-system
variation in the satellite LF. We have estimated the intrinsic rms
scatter about the mean LF using the method detailed in the Appendix.
We indicate this range with the blue error bars on the cumulative LF in
the lower panel of Fig.~\ref{fig:mw_lf} and the mean plus the rms
of the differential LF by the blue dashed line in the upper
panel. Since even in the cumulative LF, the mean number of satellites
per primary is low it is inevitable that the width of the distribution
includes zero satellites. The wide scatter illustrates the danger of
just using the MW+M31 to constrain models.
 
It is also interesting to see how the satellite luminosity function
depends on the colour and morphology of the primary galaxy.
Fig.~\ref{fig:type_lf} shows the resulting satellite LFs when
primaries of $V$-band magnitude $-21.25\pm 0.5$ are split by colour and
by concentration. In the upper panel we divide the primary galaxies 
into ``red" and ``blue" subsamples according to the well-known
colour bimodality in the colour-magnitude plane
\citep[e.g.][]{str01,bal04,zeh05}.  Following \cite{zeh05}, we use an
equivalent colour criterion of $^{0.0}(g-r)_{\rm{cut}}=0.19-0.24M_r$
(not identical to \citeauthor{zeh05} as our magnitudes are K-corrected
to $z=0.0$ rather than $z=0.1$).  We see that in this bright satellite
regime, the LF around blue primaries is lower than the LF around red
primaries.  This difference might simply reflect the relative mass of the
halos. Assuming stellar mass to correlate with halo mass we would
expect that at a fixed $V$-band magnitude blue star forming galaxies
would be less massive than their red counterparts.

The lower panel splits the sample into early and late type
where the early type is defined as having a concentration index $c \ge
2.6$. This division roughly separates 
early-type (E/S0) galaxies from late-type (Sa/b/c, Irr)
galaxies \citep{shi01}.  We see that the satellite LF of late types
is suppressed with respect to that of the early types. Given the 
well known correlation between colour and morphology this result
is consistent with the division by colour.

We can also use the colour information available in SDSS to probe the
properties of the satellites. For two bins of $V$-band primary
magnitude, Fig.~\ref{fig:sat_color} shows their satellite luminosity
functions split into red and blue subsamples using the same cut in the
colour magnitude plane as before. We see that at all but the brightest
magnitudes the satellites are predominately blue and star
forming. This is in stark contrast with the satellites in groups and
clusters where the brightest tend to be red and dead while the
faintest are blue \citep{ski09}. We also note that the LF of the red
satellites is far from a power law. It has a distinct dip in the range
from $3.0<\Delta M<5.0$ and, for the brighter primaries, the peak
$\Delta M\approx2.0$ that we noted earlier in the total LFs is clearly
present in the red subsample (and also in the blue subsample).

\begin{figure}
  \includegraphics[width=84mm]{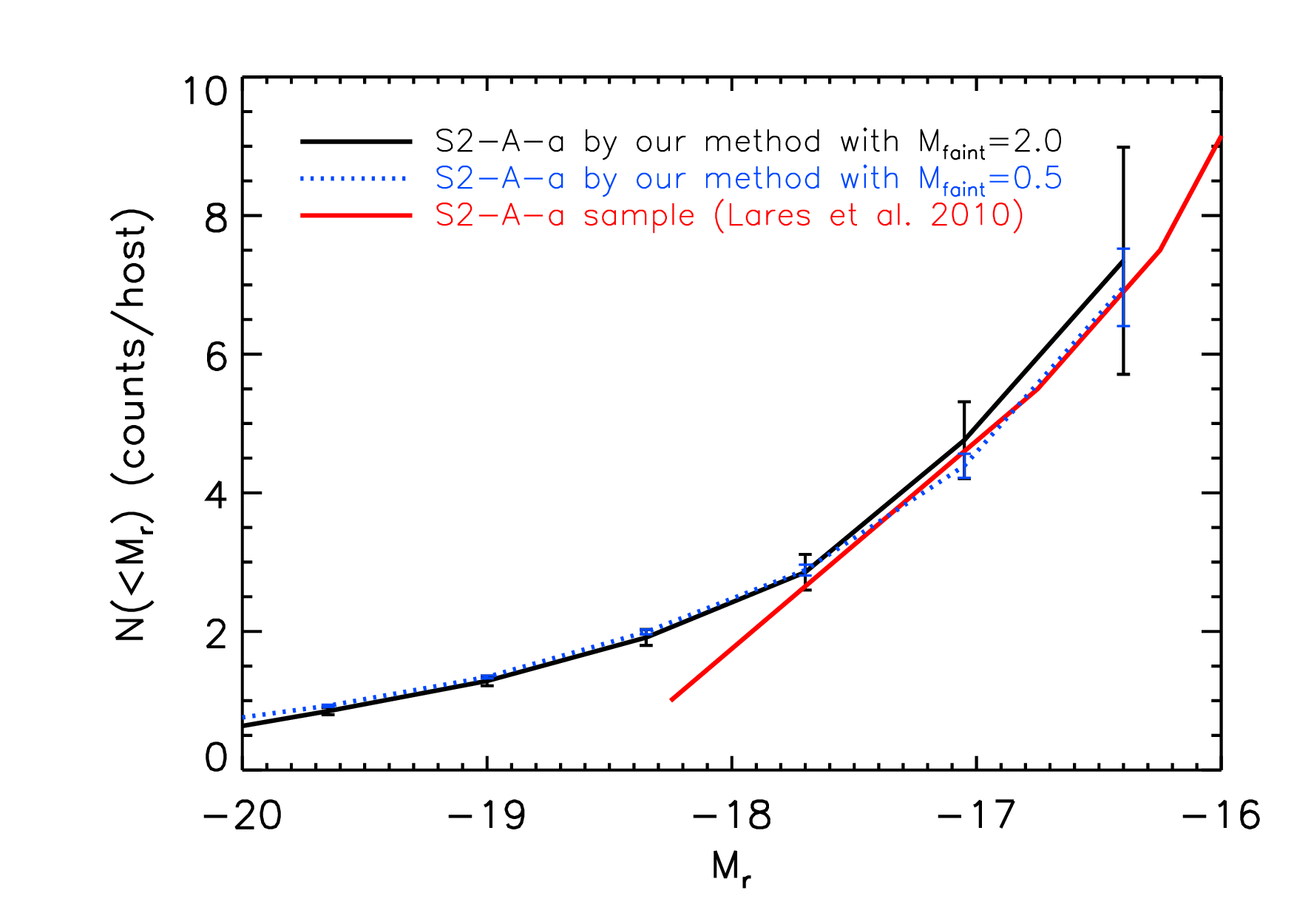}
   \caption{ The red line shows the satellite LF estimated by
             \citep{lar10} for their S0-A-a subsample.
             The solid black and dotted blue lines show our result
             for a similar sample primary galaxies, those
             brighter than -21.5 in $r$-band  
             and in the redshift range $0.03<z<0.1$, using
             $\Delta M_{\rm faint}=2.0$ and~$0.5$ respectively. 
          }
   \label{fig:comp}
\end{figure}

\section{Discussion}
\label{sec:con}

We have constructed a large sample of isolated primary galaxies and
their fainter neighbours using both the SDSS DR7 spectroscopic and
photometric galaxy catalogues. The samples are sufficiently large that
we are able to stack the systems and accurately subtract the local
background to estimate the mean satellite luminosity function (LF) and
its dependence on the luminosity, colour and morphology (optical
concentration) of the primary. Our main conclusions are:

\begin{enumerate}

\item The satellite LF is well determined over a range extending to
approximately 8 magnitudes fainter than the primary, for primaries with
V magnitudes in the range -20 to -23.

\item The satellite LF does not have a Schechter form. 
After a steep decline at the faintest magnitudes, the LF roughly
follows a fairly flat power law but there is a bump at relative
magnitude $\Delta M \simeq 2$ which is particularly significant for
brighter primaries (see Fig.~7).

\item Over the range $-14> M_{\rm V}>-19$, the mean satellite LF
around primaries of $M_{\rm V}=-21.25$ has a similar slope, but about
a factor of two lower amplitude than the average of the combined MW
and M31 LFs (see Fig.~8).

\item The amplitude of the satellite LF increases with the
luminosity of the primary. Over most of the range sampled, the
increase is approximately a factor of 2 per primary V magnitude, but
there are significant variations in the shape of the function for
primaries of different luminosity (see Fig.~7).

\item The amplitude of the satellite LF also varies with the colour
and the morphological type of the primary. Red primaries have more
satellites than blue primaries and early-type primaries have more
satellites that late-type primaries (see Fig.~9).

\item Except for the brightest objects, satellite galaxies are
predominantly blue and star-forming (see Fig.~10).

\end{enumerate}

As we were completing this work two related studies were published,
both using the SDSS DR7. \citet{liu10} used similar selection criteria
to ours to construct a sample of Milky Way-like primaries and
deconvolved for the variation of the background to determine the
frequency at which these Milky-Way like systems host satellites as
bright as the SMC and LMC. They find that 11.6\% host one such satellite
and only 3.5\% host two. And they find a mean of 0.29 satellites per primary.
This is in excellent agreement with the mean of 0.30 that we find for
satellites between 2 and 4 magnitudes (the range used by
\citeauthor{liu10}) fainter than primaries with the magnitude,
$M_{\rm V}=-20.9$, adopted by \citeauthor{liu10} For the fiducial
``Milky Way'' luminosity we have adopted here, 
$M_{\rm V}=-21.25\pm 0.5$, we find a slighly larger mean of 0.47
Magellanic cloud type satellites per primary. 

In a separate study, \citet{lar10} estimated cumulative satellite
luminosity functions and radial density profiles of satellite systems
around primaries brighter than $M_r=-20.5$. When we reproduce the
selection criteria of one of their samples using our catalogue, we
find excellent agreement for satellite magnitudes fainter than
$M_r=-18.5$, but at brighter magnitudes we find a significant excess
compared to their estimate (see Fig.~\ref{fig:comp}). This excess is
robust to changes in the value of the isolation parameter, $\Delta
M_{\rm faint}$, that we have used. 

The satellite LF probes the smallest scales visible today in the
hierarchy of galaxy formation. This statistic provides a strong test
of the \lcdm\ cosmological model, which robustly predicts the number
of subhalos that could host satellite galaxies, and a test of galaxy
formation theory, which determines which of these subhalos are
populated by visible satellites. The results so far are
encouraging. For example, the original \lcdm\ galaxy formation model
of \citet{ben02} (which predicted the population of ultrafaint
satellites subsequently discovered in the SDSS), as well as the more
recent model of \citet{qi10} predict that bright satellites like the
LMC and the SMC should be rare. This feature appeared to be a
shortcoming of the model when data were available only for the MW 
\citep{kop08}. The new results for large samples of MW-like galaxies by
\citet{liu10} and ourselves now suggest that the MW is unusual in having
such bright satellites.

According to standard theory, the satellite LF is established by
processes that regulate star formation in small halos, namely
photoionization of the gas at high redshift and supernova feedback,
acting on a population of dark matter subhalos, itself the result of
dynamical evolution from a spectrum of primoridal \lcdm\ density
perturbations. Our analysis and those by \cite{liu10} and \cite{lar10}
reveal features in the satellite LF and systematic trends with the
properties of the central galaxies. These properties encode
information about galaxy formation processes that will help develop
increasingly refined theoretical models.

\section*{Acknowledgements}
 
We thank Peder Norberg for supplying the mask and software for
quantifying the sky coverage of the SDSS DR7. We thank the
referee, Diego G. Lambas, for helpful criticism and suggestions. 
We also thank John Lucey and Tom Theuns for valuable suggestions. 
QG acknowledges a fellowship from the European Commission's Framework
Programme 7,
through the Marie Curie Initial Training Network CosmoComp
(PITN-GA-2009-238356), SMC acknowledges a Leverhulme Research
Fellowship. CSF acknowledges a Royal Society Wolfson Research Merit
Award and ERC Advanced Investigator grant 267291 COSMIWAY. This work
was supported in part by an STFC rolling grant to the Institute for
Computational Cosmology of Durham University. \appendix

\section{Estimate of the population variance}
\label{appendix:a}

\begin{figure}
 \centering \includegraphics[width=50mm]{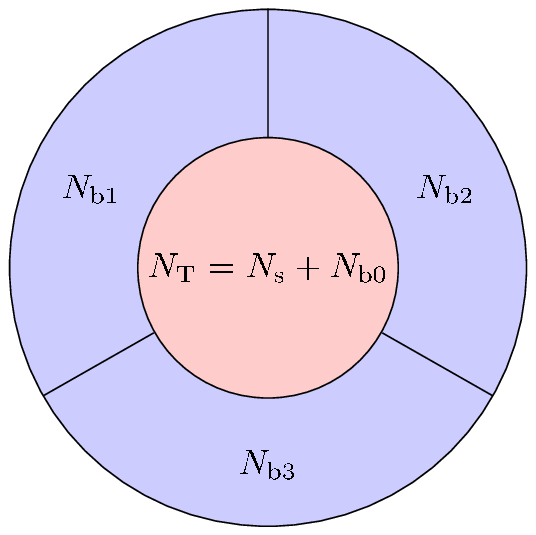} 

\caption{Schematic of the definition of quantities used in the estimation of the
 population variance. The number of genuine satellites is $\gns$ and
 the number of contaminating background galaxies in the inner area is
 $\gnbz$; the number of background galaxies in three equal area
 regions of the outer annulus are $\gmnbv$, with $i=1,2$ or $3$.  } 
 \label{fig:comp}
\end{figure}

  Here we describe the method by which we estimate the intrinsic
  variance in the number of satellites per primary. As illustrated in
  Fig.~\ref{fig:comp} we are only able directly to count the total
  number of galaxies, \begin{equation} \gnt=\gns+\gnbz,
  \label{eqn:start} \end{equation} where $\gns$ is the number of
  genuine satellites and $\gnbz$ is the number of contaminating
  background galaxies in the inner area. These two contributions
  cannot be measured separately, but we can estimate the mean number
  of satellites as,
\begin{eqnarray}
 \gmns &=&\gmnt-\gmnbz  \\
               &=&\gmnt-f\gmnout,
\end{eqnarray}
where $\gnout= \gnbi + \gnbii +\gnbiii$ is the total number of 
background galaxies in the outer area and $f$ is the ratio of the
inner to outer areas. Below we will take $f=1/3$ which is the
case for our default choice of $R_{\rm outer} =2 R_{\rm inner}$.

Here we are interested in calculating the variance in the number of
satellites, $\langle (\gns-\gmns)^2\rangle$. Starting with
equation~(\ref{eqn:start}) we can write the variance in the total
number of galaxies as
\begin{eqnarray}
\langle (\gnt-\gmnt)^2\rangle=\langle(\gns +\gnbz -
\langle(\gns+\gnbz)\rangle)^2\rangle  \nonumber \\
=\langle (\gns-\gmns)^2\rangle +
 \langle(\gnbz-\gmnbz)^2\rangle  \nonumber \\ 
 +2\langle(\gns-\gmns)(\gnbz-\gmnbz)\rangle.
\label{eqn:imp}
\end{eqnarray}
If we assume that the number of actual satellites, 
$N_{\rm s}$, around each primary is uncorrelated with the number of
background galaxies, $N_{\rm b0}$, the final cross term vanishes
to leave 
\begin{equation}
\langle (\gns-\gmns)^2\rangle=\langle (\gnt-\gmnt)^2\rangle-
\langle(\gnbz-\gmnbz)^2\rangle.
\end{equation}
The term, $\langle(\gnbz-\gmnbz)^2\rangle$ cannot be directly
measured, but, to a good approximation, we would expect it to equal the
variances, $\langle(\gnbv-\gmnbv)^2\rangle$, of each of the equal area
portions of the outer annulus. Hence, our final estimate of the
variance in the number of genuine satellites per primary can be
written as
\begin{equation}
\langle (\gns-\gmns)^2\rangle
=\langle (\gnt-\gmnt)^2\rangle-
\frac{1}{3}\sum_{i=1}^{3}\langle(\gnbv-\gmnbv)^2\rangle .
\end{equation}
For a given selection of primaries and choice of satellite absolute
magnitude, these terms will depend on the redshift of the primary. We
find a smooth variation with redshift bin and weight the variances
according to the contribution each redshift makes to the overall
estimate of the satellite luminosity function to estimate the overall
variance on the luminosity function. The result is shown in Fig.~8.

\label{lastpage}

\end{document}